\documentclass[12pt]{article}
\usepackage{latexsym,amssymb,amstext,amsmath,empheq,array}
\usepackage{slashed}
\usepackage{mathrsfs}
\usepackage{hyperref}
\usepackage{mathabx}
\usepackage{dsfont}
\usepackage{xcolor}

\renewcommand{\Large}{\large} 

\allowdisplaybreaks

\numberwithin{equation}{section}
\newcommand{\Comment}[1]{}
\newcolumntype{P}[1]{>{\centering\arraybackslash}p{#1}}

\newcommand {\cO}{{\cal O}}

\newcommand {\cQ}{{\cal Q}}
\newcommand {\cR}{{\cal R}}
\newcommand {\cS}{{\cal S}}

\def\a{\alpha}
\def\b{\beta}

\def\d{\delta}
\def\e{\epsilon}

\def\g{\gamma}

\newcommand{\cbar}{\bar{c}}
\newcommand{\Lbar}{\bar{L}}
\newcommand{\Qbar}{\bar{Q}}
\newcommand{\Rbar}{\bar{R}}

\newcommand{\be}{\begin{equation}}
\newcommand{\ee}{\end{equation}}
\newcommand{\bea}{\begin{eqnarray}}
\newcommand{\eea}{\end{eqnarray}}

\newcommand{\ba}{\begin{array}}
\newcommand{\ea}{\end{array}}

\def\double #1{#1{\hbox{\kern-2pt $#1$}}}

\newcommand{\bsubeq}{\begin{subequations}}
\newcommand{\esubeq}{\end{subequations}}

\newcommand{\chb}{{\mathfrak b}}
\newcommand{\chc}{{\mathfrak c}}

\newcommand{\eref}[1]{Eq.\,(\ref{#1})}

\begin{document}

\begin{titlepage}

\begin{center}

\vskip .3in \noindent

{\Large \bf{Extended Supersymmetric BMS$_3$ algebras\\[1mm] and Their Free
    Field Realisations 
}}

\bigskip

	{Nabamita Banerjee}$^{\,a,}$\footnote{nabamita@iiserpune.ac.in}, {Dileep P. Jatkar}$^{\,b,c,}$\footnote{dileep@hri.res.in}, {Ivano Lodato}$^{\,a,}$\footnote{ivano@iiserpune.ac.in},\\[2mm]
	 {Sunil Mukhi}$^{\,a,}$\footnote{sunil.mukhi@iiserpune.ac.in}, {Turmoli Neogi}$^{\,a,}$\footnote{turmoli.neogi@students.iiserpune.ac.in}\\

       \bigskip
       $^{a}$ \em Indian Institute of Science Education and Research,\\ Homi Bhabha Road, Pashan, Pune 411 008, India \\
       \medskip
	   $^{b}$ Harish-Chandra Research Institute,\\ Chhatnag Road, Jhusi, Allahabad, India 211019 \\
       \medskip
       $^c$ Homi Bhabha National Institute\\
Training School Complex, Anushakti Nagar, Mumbai 400085, India

       \vskip .5in
       {\bf Abstract }
       \vskip .2in
       \end{center}
       We study $N=(2,4,8)$ supersymmetric extensions of the three dimensional BMS algebra (BMS$_3$) with most generic possible central extensions.  We find that $N$-extended supersymmetric BMS$_3$ algebras can be derived by a suitable contraction of two copies of the extended superconformal algebras.  Extended algebras from all the consistent contractions are obtained by scaling left-moving and right-moving supersymmetry generators symmetrically, while Virasoro and R-symmetry generators are scaled asymmetrically. On the way, we find that the BMS/GCA correspondence does not in general hold for supersymmetric systems. Using the $\beta$-$\gamma$ and the ${\mathfrak b}$-${\mathfrak c}$ systems, we construct free field realisations of all the extended super-BMS$_3$ algebras.


\vfill
\eject

\end{titlepage}

\tableofcontents
\section{Introduction}

In the last few years there has been a renewed interest in the infinite dimensional algebra of boundary-condition-preserving symmetries of asymptotically flat spacetimes at null infinity, originally discovered by Bondi, van der Burg, Metzner, Sachs (BMS) more than half a century ago \cite{Bondi:1962px,Sachs:1962wk}.   

These (global) symmetries, originally discovered in the four dimensional context, consist of a semi-direct product of an infinite dimensional group of angle dependent translations along the null time coordinate and the Lorentz group. A local version of the BMS symmetry can also be obtained if one allows the Killing vector at asymptotic infinity of the spacetime to be finitely singular, {\em i.e.}, to be  a meromorphic function: in this case the BMS group consists of a  product of supertranslations and another infinite dimensional group of symmetries called super rotations \cite{Barnich:2009se,Barnich:2011ct}. In either case, BMS symmetries contain the usual global Poincar\'e and Lorentz group as a normal subgroup. This observation will be useful in what follows.

 It is important to stress that BMS symmetries, in their global or local form, are not isometries of flat spacetimes at infinity.  A flat solution is transformed into another flat solution containing soft gravitational hair (Goldstone bosons arising from the broken global invariance of the vacuum). Since the BMS algebra is the semi-direct product of two infinite dimensional algebras, the amount of soft hair that can be produced is also infinite. 

The seminal works of Strominger et al. \cite{Strominger:2013jfa,He:2014laa,He:2014cra,Strominger:2013lka} shed some light on the connection between BMS symmetries and the soft graviton (and photon) theorems proved by Weinberg \cite{Weinberg:1965nx}.  They showed that a diagonal subgroup of the BMS symmetries defined on $\mathcal{I}^-$ and $\mathcal{I}^+$ is an exact symmetry of the S-matrix of a gravitational (or gauge) theory. This has led to an improved understanding of the relevance of BMS symmetries, and it is believed that the analysis of boundary-conditions-preserving symmetries on a null surface might have many more surprises in store. 

Quite recently, for instance, there have been several papers aimed at improving our understanding of gravitational waves and their scattering amplitudes, the so-called gravitational memory effect \cite{Strominger:2014pwa,Pasterski:2015tva}, and at resolving some important open problems in physics, such as the information paradox \cite{Hawking:2016msc}. Additionally, in \cite{Hawking:2016msc} the black hole horizon, and not just null infinity, was considered as a boundary of the spacetime, and by solving the constrained Killing equations, the presence of an infinite amount of soft hair at the horizon of a black hole was established. This possibility is quite tempting in that it seem to pave the way for a quantum description of black holes, including their microscopic and statistical entropy \cite{Averin:2016ybl,Donnay:2015abr,Afshar:2016wfy,Afshar:2016uax} without the use of supersymmetry
\footnote{Indeed, little is known about the microscopics of non-supersymmetric black holes. The only explicit results are for the statistical entropy of solutions with an AdS$_3$ factor in the near-horizon geometry  \cite{Strominger:1997eq} (see next footnote), or connected to those via dimensional reductions. Note that to obtain the matching between macroscopic and statistical entropies, the inclusion of the full effective higher derivative corrections is paramount \cite{Banerjee:2016qvj}. }.
Especially interesting are the results obtained in \cite{Donnay:2015abr,Afshar:2016wfy}: the former paper treats two simple cases of 3- and 4-dimensional black holes with certain boundary conditions at the horizon while the latter one treats a different 3-dimensional solution. Their results are similar in that they connect the statistical entropy with the zero modes of the boundary-condition-preserving symmetries at the horizons. However, the details of the computations show that starting from very specific boundary conditions, one  can be led to algebras very different from the standard BMS algebra (or its generalizations), such as the Heisenberg algebra \cite{Afshar:2016wfy}.

The results of \cite{Afshar:2016uax} are, in a sense, even stronger: among all the infinite microstates available, which preserve the boundary conditions at the horizon and at infinity, the only `physical' ones exist at the horizon but have vanishing Virasoro generators at infinity. Using the Hardy-Ramanujan expansion, the area law is obtained. It is worth mentioning that the use of conformal methods at the horizon was already proposed by Carlip \cite{Carlip:1999cy} almost a couple of decades ago, as an extension to any dimension of the famous results of Strominger valid only in three dimensions \cite{Strominger:1997eq}\footnote{Strominger obtained the entropy of asymptotically locally AdS$_3$ black holes using the Cardy formula, which computes the statistical entropy of states in a two dimensional CFT. This result was in turn obtained by exploiting the correspondence between the algebra of asymptotic charges of AdS$_3$ spacetimes and the Virasoro algebra \cite{Brown:1986nw}, which can be considered a progenitor of the AdS/CFT correspondence.}.
However, the final results were strongly dependent on a very specific set of coordinates. Later on, it was shown \cite{Guica:2008mu} that a set of boundary conditions at the horizon of a 4d Kerr black hole exists, such that the algebra of boundary-condition-preserving symmetries is a copy of the Virasoro algebra, and again the entropy was computed through the use of the Cardy formula. None of the above results, however, contained intuition about the presence of soft hair and more importantly a precise counting of microstates, necessary for a complete quantum description of black hole solutions, was lacking.

In order to make contact with the all-orders (not just leading-order) microscopic counting results in the literature, which rely on supersymmetry, we would like to investigate supersymmetric extensions of BMS symmetries. 
 The BPS black hole degeneracy could then be compared with the counting of soft hair of the extended supersymmetric BMS algebra.  Although ideally we would like to carry out this comparison in four dimensions, in the present paper we will consider examples of extended $N=(2,4,8)$ supersymmetric BMS$_3$ algebras.

The $N=1$ supersymmetric extension of BMS$_4$ was already obtained many years ago \cite{Grisaru:1977kk,Awada:1985by} and more recently, in the context of string theory, in \cite{Avery:2015iix}.
In a recent paper Barnich et al. \cite{Barnich:2014cwa} addressed the problem in 3 dimensions and, using a set of very specific boundary conditions, obtained the $N=1$ super BMS$_3$ charge algebra as a flat limit of the algebra of charges of asymptotically AdS$_3$ spaces, namely the super-Virasoro algebra.
 We follow a similar procedure and obtain $N$-extended ($N=2,4,8$) super-BMS$_3$ algebras. To find the superalgebras we must choose the flat limit to be such that the anticommutator of two supersymmetries closes into a translation, given the fact that a subgroup of the full super-BMS algebras must be equivalent to the physical super Poincar\'e algebra. Additionally, different supercharges should anti-commute. Thus we require: $\{Q^I,(Q^J)^\dagger\}=\delta^{IJ}\,P$.

The precise scaling of super-Virasoro generators required to obtain a super-BMS algebra admits some ambiguities, and we will attempt to resolve these in the present paper. For example in the $N=1$ supersymmetric case, the original superconformal algebra only has a single chiral supercharge and therefore it has to be scaled by itself. However once we start with at least $N=(1,1)$ supersymmetry, one encounters the possibly of either scaling them independently and symmetrically, or else taking linear combinations of left and right-moving supercharges and scaling them asymmetrically, as is done with Virasoro generators. The same question arises for R-symmetry generators. 

We will find that in fact all bosonic generators must be scaled in an asymmetric way, while the supersymmetry generators need to be scaled in a symmetric fashion. This has a consequence for the proposed relation between the BMS and Galilean Conformal Algebras \cite{Bagchi:2010zz}. As we will see, in the presence of supersymmetry this correspondence no longer holds. Another important observation is that the BMS algebra which we find as the In\"on\"u-Wigner
contraction of various superconformal algebras, is not the most generic centrally extended algebra. This is because the relation among different central charges in a superconformal theory is ``broken'' by the contraction process. We will address this point in section 4. Finally, we will provide the free-field realizations of our various extended super-BMS$_3$ algebras, as was done recently for the $N=1$ super-BMS algebra\cite{Banerjee:2015kcx}.

The paper is organized as follows: in Section \ref{sec:2} we show explicitly that the Poincar\'e algebra in 3 dimensions is a subalgebra of BMS$_3$. This result will be important in the next sections, since the extended super-BMS$_3$ algebras we find must contain, as a subalgebra, the super-Poincar\'e algebra in 3 dimensions. In Sections \ref{sec:3},\ref{sec:4},\ref{sec:5} we construct the possible $N=2,4,8$ super-BMS$_3$ algebras arising from consistent contractions of the asymptotic symmetry algebra, isomorphic to the extended super-Virasoro algebra. In each of the sections we also present the free field realizations for each case, via the usual bosonic $\b-\g$ system and the addition of Grassmann-odd ghost ${\mathfrak b}-{\mathfrak c}$ systems relevant to the supersymmetric extensions. While the choice of the contraction for the minimally extended case is straighforwardly obtained in analogy to the $N=1$ case, the $N=4,8$ cases need more attention, as R-symmetry generators will in general be present. For this reason, section \ref{sec:4} treats in detail the available scalings for the R-currents, out of which the asymmetric scaling is singled out as the only one which allows R-symmetry to be present in the final algebra. This hints at which scaling must be used also in the maximally extended case treated in Section \ref{sec:5} to obtain a proper $N=8$ super-BMS$_3$ algebra. In fact for $N=8$, symmetric scalings for the R-currents give rise to an inconsistent contraction, which confirms the $N=4$ observation. In hindsight, by a reverse argument we could have started from the $N=8$ case, where only the asymmetric contraction for the R-currents is consistent, and used it to discover the $N=4$ case as a subalgebra. Finally, we present a summary of the results in section \ref{sec:6}. We also have two appendices: in the first one we present \emph{all} other possibilities for the well defined contraction of the supersymmetry generators in the $N=2$ case, and show that the BMS$_3$/GCA$_2$ isomorphism, proposed in \cite{Bagchi:2010zz} does not hold at the supersymmetric level \footnote{An analogous observation has been made by analyzing the representation theory of both algebras \cite{Campoleoni:2015qrh} and in the context of (non-supersymmetric) higher spin symmetries \cite{Campoleoni:2016vsh}. Furthermore the $N=2$ BMS$_3$ algebra we will present was already derived in the context of the two dimensional null string and spinning ambitwistor string \cite{Casali:2016atr}. We thanks Max Riegler and Piotr Tourkine for bringing these references to our attention. }. The second appendix  shows the existence of two inequivalent $N=2$ super-BMS$_3$ algebras, arising from contractions respectively of the $(2,0)$ and $(1,1)$ Virasoro algbra. In principle, such arguments can also be generalized to the $N=4$ case where, for instance, inequivalent contractions of the $(2,2)$ and $(4,0)$ cases can be considered.

\section{Poincar\'e as a subgroup of BMS$_3$ in flat Minkowski}
\label{sec:2}

We want to study the behavior of global residual supersymmetries along the null directions of the asymptotically flat space-time. This will be obtained by carrying out suitable scalings on the asymptotic symmetries for asymptotically AdS space-times, namely superconformal symmetries. In the bosonic case, there is more than one way to perform such scalings on combined left- and right-moving generators. One way of doing so violates the mode number, while the other way preserves it. This in particular manifests itself in the isomorphism between the BMS$_3$ and Galilean Conformal (GCA$_2$) algebras\cite{Bagchi:2010zz}. For supersymmetry one may anticipate that something similar should hold. However, as we will soon see, this does not happen. In fact, one can start from physical constraints to select the correct BMS algebra: we impose that the 
anticommutators of a supercharge with its Hermitian conjugate must be proportional to a translation generator with a positive coefficient.
\begin{equation}
\label{eq:phys_constr}
\{Q^I,(Q^I)^\dagger\}\sim P \;
\end{equation} 
As we will show, this constraint is sufficient to uniquely determine \emph{all} the supersymmetric-extended BMS$_3$ algebras and implies that they differ from the super-GCA$_2$ algebras.

Before treating the supersymmetric extensions of BMS we will need to identify the Poincar\'e subgroup inside the BMS$_3$ group. Although this has already been stated in some earlier works (\cite{Barnich:2013yka}), we present here a  
systematic procedure to arrive at the Poincar\'e subgroup. Eventually, we shall extend this procedure to supersymmetric cases. Let us start with a generic metric in the BMS gauge (which includes Bondi coordinates)\cite{Barnich:2010eb}:
\begin{align}
ds^2&=e^{2\,\beta}\,\frac{V}{r}{\rm d}u^2-2\,e^{2\beta}\,{\rm d}u\,{\rm d}r+r^2\,e^{2\varphi}\,({\rm d}\phi-U\,{\rm d}u)^2\;,
\nonumber\\
g_{\mu\nu}&=\begin{pmatrix}e^{2\,\beta}\,\frac{V}{r} & -e^{2\beta} & -r^2\,e^{2\varphi}\,U
\\
-e^{2\beta} & 0 &0
\\
-r^2\,e^{2\varphi}\,U & 0 & r^2\,e^{2\varphi} 
\end{pmatrix}
\end{align}
We will initially focus on flat Minkowski spacetime, which is given by $\varphi=\beta=U=0$ and $V=-r$. In the general case, which we will analyze later, each of these functions has a non-trivial fall-off condition as we approach null infinity. We start with flat space in 3D parametrised as:
\begin{equation}
ds^2=-{\rm d}t^2+{\rm d}x^2+{\rm d}y^2\;,
\end{equation}
with $x=r\,\cos\phi$, $y=r\,\sin\phi$, $u=t-r$ , $r=\sqrt{x^2+y^2}$. We want to understand how a rigid translation of all the coordinates $(t,x,y)$ is represented in the BMS frame, specifically on the $u$-coordinate. 

Let us assume that under rigid translations $t\rightarrow t^\prime=t+a$, $x\rightarrow x^\prime=x+b$ and $y\rightarrow y^\prime=y+c$. This means that:
\begin{align}
u^\prime&=t^\prime-r^\prime=t+a-\sqrt{(x+b)^2+(y+c)^2}
\nonumber\\
&\sim t+a-r(1+\frac{1}{r}(b\,\cos\phi+c\sin\phi)+\cO(b^2,c^2))
\nonumber\\
&=u+a-b\,\cos\phi-c\sin\phi+\cO(b^2,c^2)
\nonumber\\
&=u+a-\frac{1}{2}(b-i\,c)\, e^{i\,\phi}-\frac{1}{2}(b+i\,c)e^{-i\,\phi}+\cO(b^2,c^2)
\end{align} 
where, instead of expanding in large $r$, we have expanded in the small translation parameters. This way we can extend our definitions to the bulk, where $r$ takes a finite value. As is easily seen, the translations of the coordinates $(t,x,y)$ correspond to a supertranslation of the $u$ coordinate, parametrized by the basis vectors $e^{i\,n\,\phi}$ with $n=-1,0,1$ and linear combinations thereof. The coefficients of these basis vectors give us the translation parameters.

Note that in general such transformation will change the coordinates $r$ (obviously) and $\phi$ (less obvious but still true), i.e.
\begin{equation}
\tan\phi^\prime=\frac{y^\prime}{x^\prime}=\frac{y+c}{x+b}=\frac{y}{x}\big(1-\frac{b}{x}+\frac{c}{y}\big)=\tan\phi-\frac{1}{x}\big(b\,\tan\phi-c\big).
\end{equation}
The last terms, expanded in terms of the small translation parameters could be equivalently expanded in terms of large $r$ and immediately one notices that the change in $\phi$ is suppressed in powers of $1/x\sim1/r$.

Thus we have identified the generators of the translation subgroup of BMS. Next we want to identify the 3 generators of Lorentz transformations (1 rotations, 2 boosts). To do so we look at the complete form of the Killing vectors of the above metric, with fall-off conditions fixed by the requirement of the BMS gauge. The most general result for the Killing vectors, which can be extended also to the bulk \cite{Barnich:2010eb}, is given by (the form of the metric is invariant up to subleading terms in $r$) :
\begin{empheq}[left=\empheqlbrace]{align}
&\xi^u=f\,\;, \qquad f=e^\varphi\Big(T(\phi)+\int_0^u\,du^\prime\,e^{-\varphi}(\partial_\phi\,Y(\phi)+Y(\phi)\,\partial_\phi\,\varphi)\Big)
\nonumber\\
&\xi^\phi=Y(\phi)-e^{-2\,\varphi}\,\partial_\phi\,f\,\int_r^\infty\,dr^\prime\,{r^\prime}^{-2}\,e^{2\,\beta}
\nonumber\\
&\xi^r=-r\Big(\partial_\phi\,\xi^\phi-\partial_\phi f\,U+\xi^\phi\,\partial_\phi\,\varphi+f\,\partial_u \varphi\Big)
\end{empheq}
Again, for the specific example of Minkowski, the above components simplify greatly, giving:
\begin{empheq}[left=\empheqlbrace]{align}
&\xi^u=f\,\;, \qquad f=T(\phi)+u\,\partial_\phi\,Y(\phi)
\nonumber\\
&\xi^\phi=Y(\phi)-\frac{1}{r}\,\partial_\phi\,f\,
\nonumber\\
&\xi^r=-r\,\partial_\phi\,\xi^\phi
\end{empheq}
Note that $T$ and $Y$ are integration constants.
At null infinity, {\em i.e.}, when $r\rightarrow \infty$, these vectors become the Killing
vectors, and are simplified further. 

Let us see the action of
these vector fields on the metric.
First of all, we write down the decomposition in terms of a basis
dictated by the angular coordinate $\phi$. One can take $e^{i n\phi}$ to generate
 the supertranslations and, at the same time, the conformal reparametrizations of 
the circle which are, in fact, Lorentz transformations. Hence we have: 
\begin{equation}
\bar\xi_{T,Y}=\big(T(\phi)+u\,\partial_\phi\,Y(\phi)\big)\partial_u+\big(Y(\phi)-\frac{1}{r}\,\partial_\phi\,f\big)\partial_\phi-r\,\partial_\phi\,\xi^\phi\,\partial_r
\end{equation}
Let us now take the limit $r\rightarrow \infty$ and obtain:
\begin{equation}
\bar\xi_{T,Y}=\big(T(\phi)+u\,\partial_\phi\,Y(\phi)\big)\partial_u+Y(\phi)\partial_\phi-r\,\partial_\phi\,Y(\phi)\,\partial_r
\end{equation}
where the last term can be dropped, because, in taking the limit $r\rightarrow \infty$, no function or field will depend on $r$ anymore, or equivalently, we can take a surface $r={\rm const}$ and $r$
very large and see the action of the vector $\bar\xi$ on the metric defined on that surface, etc. As we anticipated the Killing vectors can be decomposed in terms of a basis, as follows:
\begin{align}
J_n&=\{\bar\xi_{T,Y}: T=0\,,\quad Y=e^{i\,n\,\phi}\}
\nonumber\\
P_n&=\{\bar\xi_{T,Y}: T=e^{i\,n\,\phi}\,,\quad Y=0\}
\end{align}
We have identified the 3 generators $t_n$ which give proper
translations in spacetime. 

The next step is to identify the Lorentz
transformations. Note that there exists one such Lorentz
transformation which will rotate the circle but does not change the
coordinate $u$. This corresponds in fact to the rotations of the
circle, which is trivially given by $J_0$, which gives
$\bar\xi=\partial_\phi$ as usual. The other two Lorentz boosts will
mix the coordinates $\phi$ with $t$ and $r$, or $u$ and $r$, in
general. Those are given by $J_{1}$ and $J_{-1}$, as it is easy
to check, by writing down the algebra relations of these 3 generators
and see that it closes into itself. Furthermore, if we now add $P_0$,
$P_{-1}$ and $P_1$ to the mix we obtain exactly the Poincar\'e
subalgebra at infinity\footnote{Which, up to some field dependent terms, can be extended to
the null generators in the bulk of Minkowski spacetime (the fields $\phi$, $U$, etc will still remain fixed, only the $r-$dependence will change, making the explicit form of the vector field $\bar\xi$ slightly more complicated).}.  Thus, we explicitly understand how global Poincar\'e is a subalgebra of BMS$_3$ algebra.

In following sections, we will  present supersymmetric extensions of BMS$_3$ algebra. These algebras are not so well-understood physically. Our guideline to write to such algebras will be to: 1) define a proper contraction to get a finite algebra and 2) identify the global super-Poincar\'e sub-algebra inside them. As we shall see below, contraction can be carried out in different ways leading to different algebras, but some of them do not contain the proper super-Poincar\'e algebra and therefore can be rejected on physical grounds.

\section{$N=2$ super-BMS$_3$ algebra }
\label{sec:3}

We can now start to write down extended super-BMS$_3$ algebras. Three-dimensional $N=1$ super-BMS has already been presented in \cite{Barnich:2015sca} and a free-field realization of the same was given in \cite{Barnich:2014cwa}.  Here, we first extend the BMS algebra in 3d to the super-BMS algebra with $N=2$. We also provide a free-field realization of this extended algebra. 

\subsection{Construction of $N=2$ super-BMS algebra}

Let us start from the $N=1$ case dealt with in \cite{Barnich:2014cwa,Barnich:2015sca}, in connection to gauged supergravity. In \cite{Barnich:2014cwa} the algebra of asymptotically AdS$_3$ backgrounds closes into two copies of the Virasoro algebra, one of which is augmented by supersymmetry. This corresponds to $(1,0)$ supersymmetry. Specifically, the algebra reads (here we use commutators, instead of Poisson brackets):
\begin{align}
 [L_m,L_n]&=(m-n)\,L_{m+n}+\frac{c}{12}\,m(m^2-1)\,\delta_{m+n,0}\;, &[L_m,Q_r]=\big(\tfrac{m}2-r\big)\,Q_{m+r}\;,
\nonumber\\
[\Lbar_m,\Lbar_n]&=(m-n)\,\Lbar_{m+n}+\frac{\cbar}{12}\,m(m^2-1)\,\delta_{m+n,0}\;,
&\{Q_r,Q_s\}=\,L_{r+s}+\frac{c}{6}\,r^2\,\delta_{r+s,0}\;.
\end{align}
Note that there are no $\Qbar$ generators in this system.

From this, by performing a suitable contraction one obtains the asymptotic algebra of $N=1$ supergravity on asymptotically flat space. This corresponds to the minimal supersymmetrization of the BMS$_3$ algebra. To perform the contraction, we need to redefine our generators and take the (flat-space) limit $L\rightarrow \infty$ or, equivalently $\e=\frac{1}{L}\rightarrow 0$:
\begin{align}
\label{eq:contraction_1}
P_m&=\lim_{\e\rightarrow 0}\,\e \,(L_m+\Lbar_{-m})\;,\quad J_m=\lim_{\e\rightarrow 0}\, (L_m-\Lbar_{-m})\;,\quad \cQ_r=\lim_{\e\rightarrow 0}\,\sqrt{\e}\,Q_r\;,
\nonumber\\
c_1&=\lim_{\e\rightarrow 0}\,(c-\cbar)\;,\qquad\qquad c_2=\lim_{\e\rightarrow 0}\,\e\,(c+\cbar)\;.
\end{align}
As is well-known, the left-and right-moving modes of the original pair of Virasoro algebras need to be combined with equal and opposite indices and the two combinations scaled asymmetrically, as above, in order to get the BMS algebra. In the limit, the algebra for $P_m$, $J_m$, and $Q_r$ reads
\footnote{The notation used here differs from that in our last paper.  The generators $J_m,\, P_m$ here were denoted as $L_m,\, M_m$ in \cite{Banerjee:2015kcx}.}:
\begin{align}
\label{eq:algebra_gen}
[J_m,J_n]&=(m-n)\,J_{m+n}+\frac{c_1}{12}\,m(m^2-1)\,\delta_{m+n,0}\;,
\nonumber\\
[J_m,P_n]&=(m-n)\,P_{m+n}+\frac{c_2}{12}\,\,m(m^2-1)\,\delta_{m+n,0}\;,
\nonumber\\
[P_m,P_n]&=0\;,\qquad
[P_m,\cQ_r]=0\;,
\nonumber\\
[J_m,\cQ_r]&=\big(\tfrac{m}2-r\big)\,\cQ_{m+r}\;,\qquad
\{\cQ_r,\cQ_s\}=P_{r+s}+\frac{c_2}{6}\,r^2\,\d_{r+s,0}\;.
\end{align}
To keep our results completely general, we have considered the case $c\neq \cbar$. This is the $N=1$ BMS$_3$ algebra presented in \cite{Barnich:2014cwa,Barnich:2015sca}. Notice that while the Virasoro generators $L,\Lbar$ are combined into two linear combinations that are scaled asymmetrically, the supercharge is just scaled directly without taking any linear combination (there is no other choice here, since there is one holomorphic supercharge and no anti-holomorphic one). Notice also that the relative signs between the Virasoro generators $L,\Lbar$ in $P$ and $J$ are uniquely determined. Choosing the reverse of the signs above leads to a divergent limit.

Now we extend this to an $N=2$ BMS$_3$ algebra. The starting point this time will be the $(1,1)$ superconformal algebra, which is two copies of the chiral $N=1$ superconformal algebra\footnote{There is another way to get to this algebra, by starting with the $N=(2,0)$ superconformal algebra. We will analyze this construction in appendix \ref{sec:ineq-n=2-super} in detail.}:
\begin{align}
\label{eq:double_Vir}
[L_m,L_n]&=(m-n)\,L_{m+n}+\frac{c}{12}\,m(m^2-1)\,\delta_{m+n,0}\;, \nonumber \\
[\Lbar_m,\Lbar_n]&=(m-n)\,\Lbar_{m+n}+\frac{\cbar}{12}\,m(m^2-1)\,\delta_{m+n,0}\;,
\nonumber\\
[L_m,Q_r]&=\big(\tfrac{m}2-r\big)\,Q_{m+r}\;, 
\qquad\qquad\qquad\quad [\Lbar_m,\Qbar_r]=\big(\tfrac{m}2-r\big)\,\Qbar_{m+r}\;,
\nonumber\\
\{Q_r,Q_s\}&=L_{r+s}+\frac{c}{6}\,\left(r^2-\frac{1}{4}\right)\,\delta_{r+s,0}\;,
\qquad \{\Qbar_r,\Qbar_s\}=\Lbar_{r+s}+\frac{\cbar}{6}\,\left(r^2-\frac{1}{4}\right)\,\delta_{r+s,0}\;.
\end{align}
We will scale the Virasoro generators as before, but for the supercharges there seems to be more than one choice. Either one can scale both of them in the same way, or one can combine their left and right-movers and scale the linear combinations differently. For the first option, we define the scaled generators:
\begin{equation}
\label{eq:ferm_contr_1}
\cQ^1_r=\lim_{\e\rightarrow 0}\,\sqrt{\e}\,Q_{r}\;,\qquad
\cQ^2_r=\lim_{\e\rightarrow 0}\,\sqrt{\e}\,\Qbar_{-r}\;.
\end{equation}
The scaling performed in \eref{eq:ferm_contr_1} will be referred to as ``symmetric scaling'' because the factor of $\e$ is the same for both holomorphic and anti-holomorphic generators of the original superconformal algebra. Notice, however, that the re-definition is not completely symmetric since the mode number is preserved on the holomorphic side but flipped on the anti-holomorphic side.

After taking the $\epsilon\to 0$ limit one recovers the following super-BMS$_3$ algebra:
\begin{align}
\label{eq:BMS_ferm_alg}
[J_m,J_n]&=(m-n)\,J_{m+n}+\frac{c_1}{12}\,m(m^2-1)\,\delta_{m+n,0}\;,
\nonumber\\
[J_m,P_n]&=(m-n)\,P_{m+n}+\frac{c_2}{12}\,\,m(m^2-1)\,\delta_{m+n,0}\;,
\nonumber\\
[P_m,P_n]&=0\;,\qquad
[P_m,\cQ^a_r]=0\;,
\nonumber\\
[J_m,\cQ^a_r]&=\big(\tfrac{m}2-r\big)\,\cQ^a_{m+r}\;,
\nonumber\\
\{\cQ^a_r\,,\cQ^b_s\}&=\frac{1}{2}\delta^{ab}\left[P_{r+s}+\frac{c_2}{6}\,\left(r^2-\frac{1}{4}\right)\,\d_{r+s,0}\right]\;,
\end{align}
where $a,b=1,2$.  This is the most generic centrally extended three dimensional $N=2$ super BMS$_3$ algebra, that one obtains by contraction of the corresponding superconformal algebra.
We see that the translation subgroup of the BMS algebra appears on the right hand side of the anti-commutator of both supercharges $\cQ^\pm$, just as expected in supergravity. Indeed the super-Poincar\'e algebra has the following generators:
\begin{equation}
J_{\pm1}, J_0,P_{\pm1},P_0, {\cal{Q}}_{\pm\frac{1}{2}} .\nonumber
\end{equation}

The second option corresponds to ``asymmetric scaling'' where one takes linear combinations like $Q_r\pm \Qbar_{-r}$ and scales the two differently. It is easy to verify that this approach does not lead to a consistent algebra, because one anti-commutator involving the resulting supercharges turns out to be divergent. Details are provided in appendix A.

Let us recall here that in the bosonic case, the BMS algebra was found to be isomorphic to the Galilean Conformal Algebra (GCA), even though the latter involves a different contraction of the conformal generators. Specifically, in the GCA one performs asymmetric scaling of the linear combinations $L_n\pm \Lbar_n$, unlike BMS where one takes the combinations $L_n\mp \Lbar_{-n}$. Despite the different combinations, after asymmetric scaling one finds the same algebra. This was dubbed the BMS/GCA correspondence. Clearly it is of interest to know if this correspondence extends to the supersymmetric case. The supersymmetric Galilean Conformal Algebra (SGCA) with $N=2$ supersymmetry has been constructed in Ref.\cite{Mandal:2010gx} using asymmetric scaling of linear combinations of the supercharges of the form $Q_r\pm \Qbar_r$. However, as we pointed out above, asymmetric scaling of supercharges does not lead to a valid algebra in the BMS case. It follows that in the presence of extended supersymmetry, the ``BMS/GCA correspondence'' of Ref.\cite{Bagchi:2010zz} does not hold. Therefore this correspondence appears to be an accidental connection limited to the bosonic case.

\subsection{Free-field realization of $N=2$ super-BMS}

A free-field realization of the $N=1$ super-BMS$_3$ algebra was presented in Ref.\cite{Banerjee:2015kcx}. Here we present a realization of the $N=2$ algebra along those lines. We introduce a bosonic $\beta-\gamma$ system of dimensions $(2,-1)$ and two independent pairs of Grassmann-odd ${\mathfrak b}-{\mathfrak c}$ ghost systems of dimensions $(\frac{3}{2},-\frac{1}{2})$\footnote{We have chosen our notation to avoid confusion between the ghosts $\chc_1,\chc_2$ and the central charges $c_1,c_2$.}. These ghost fields satisfy the operator product expansions:
\be
  \label{eq:11a}
\gamma(z)\beta(w)\sim\frac{1}{z-w}, \quad \chb^a(z)\chc^b(w)\sim\frac{\delta^{ab}}{z-w} .\\
\ee
The free-field realization is most conveniently understood after writing the BMS algebra in terms of operators. We therefore define:
\begin{equation}
  \label{eq:1}
  \begin{split}
    T(z) &= \sum_{n\in Z} J_nz^{-n-2}\\
    P(z) &= \sum_{n\in Z} P_nz^{-n-2}\\
    \cQ^a(z) &= \sum_{r\in Z+\frac12}\cQ^a_r\, z^{-r-\frac32}
  \end{split}
\end{equation}
The OPE form of $N=2$ super-BMS$_3$ can be easily derived from \eref{eq:BMS_ferm_alg} above:
\be
  \label{eq:12}
  \begin{split}
T(z)T(w)&\sim\frac{1}{2}\frac{c_1}{(z-w)^4}+2\frac{T(w)}{(z-w)^2}+\frac{\partial T}{z-w}\\
T(z)P(w)&\sim\frac{1}{2}\frac{c_2}{(z-w)^4}+2\frac{P(w)}{(z-w)^2}+\frac{\partial P}{z-w}\\
T(z)\mathcal{Q}^a(w)&\sim\frac{\frac{3}{2}\mathcal{Q}^a(w)}{(z-w)^2}+\frac{\partial \mathcal{Q}^a}{z-w}\\
\mathcal{Q}^a(z)\mathcal{Q}^b(w)&\sim\frac{1}{3}\frac{c_2}{(z-w)^3}\delta^{ab}+\frac{P(w)}{z-w}\delta^{ab}\
\end{split}
\ee

We now choose the $N=2$ super-BMS$_3$ generators as:
\be
  \label{eq:13}
  \begin{split}
T(z)&=-\frac{3}{2}:\chb^1\partial \chc^1:(z)+\frac{1}{2}:\chc^1\partial \chb^1:(z)-\frac{3}{2}:\chb^2\partial \chc^2:(z)\\
&\qquad+\frac{1}{2}:\chc^2\partial \chb^2:(z)-2:\beta\partial\gamma:(z)-:\gamma\partial\beta:(z)-\lambda\partial^3\gamma(z)\\
P(z)&=\beta(z)\\
\mathcal{Q}^1(z)&=\chb^1(z)+\frac{1}{2}:\beta \chc^1:(z)+\lambda\,\partial^2\chc^1(z)\\
\mathcal{Q}^2(z)&=\chb^2(z)+\frac{1}{2}:\beta \chc^2(z):+\lambda\,\partial^2\chc^2(z)\
\end{split}
\ee
where $T(z)$ and $P(z)$ generate the three-dimensional BMS algebra and $Q^1$ and $Q^2$ are the supersymmetry currents. With this choice we get the desired OPE's of eq.(\ref{eq:12}).
This free field realization fixes the central charge $c_1=-4$ and $c_2=12\lambda$. The first one, $c_1$, can be made arbitrary by adding independent canonical
free fields to this system.


\section{$N=4$ super-BMS$_3$ algebra}
\label{sec:4}

To obtain a $N=4$ BMS algebra we start from a theory of gauged supergravity, as before, admitting (2,2) supersymmetry. As the previous example taught us, we need to use asymmetric scaling for the bosonic generators $L_n,\Lbar_n$ and symmetric scaling for the fermionic ones, to obtain a supersymmetric BMS algebra with the correct Poincar\'e subalgebra. The novel feature in this case is the appearance of an R-symmetry generator $R_m$ in the $N=2$ super-Virasoro algebra, under which the supercharges carry charges $\pm 1$. The algebra reads\footnote{ Note the factor in front of the central charge extension of the supercharge anti-commutator is now changed by a factor of $2/3$, compare \cite{Mandal:2010gx} with \cite{Barnich:2014cwa} and \cite{Mandal:2016lsa}.}:
\begin{align}
\label{eq:Virasoro22}
[L_m,L_n]&=(m-n)\,L_{m+n}+\frac{c}{12}\,m(m^2-1)\,\delta_{m+n,0}\;, &
[R_m,R_n]&=\frac{c}{3}\,m\,\d_{m+n,0}\;,\nonumber\\
[L_m,R_n]&=-n\,R_{m+n}\;, \quad [L_m,Q^\pm_r]=\big(\tfrac{m}2-r\big)\,Q^\pm_{m+r}\;, &
[R_m,Q^\pm_r]&=\pm Q^\pm_{r+m}\;,
\nonumber\\
\{Q^+_r,Q^-_s\}&=L_{r+s}+\tfrac12\,(r-s)\,R_{r+s}+\frac{c}{6}\,(r^2- \frac{1}{4})\,\delta_{r+s,0}\;, &
\{Q^\pm_r,Q^\pm_s\}&=0
\end{align}
together with an identical anti-holomorphic counterpart with central charge $\cbar$.  Note that $(Q^+_r)^\dagger=Q^-_{-r}$ and similarly for $\Qbar$.

In order to be completely general we will analyze both possible types of scalings (symmetric and asymmetric) for the R-currents in the following, and will find that only the asymmetric scaling leads to a consistent algebra satisfying the physical requirement \eqref{eq:phys_constr}. 

\Comment{
As it is easy to see (see appendix for more details) the mode mixing of the bosonic generators enforces mode mixing in the R-symmetry generators as well, so in either scaling we will always consider mixed modes combinations. {\color{red} What does this mean? In symmetric scaling there is no need for mode mixing.} 
}

\subsection{$N=4$ super-BMS$_3$ with asymmetric scaling for the R-currents}

To obtain the $N=4$ BMS algebra, we use the contraction \eqref{eq:contraction_1} to which we add the generalized \eqref{eq:ferm_contr_1} and the contractions for the new U$(1)$ current generators:
\begin{align}
\label{eq:contraction_1bis}
J_m&=\lim_{\e\rightarrow 0}\,(L_m-\Lbar_{-m})\;, & P_m&=\lim_{\e\rightarrow 0}\,\e (L_m+\Lbar_{-m})\;,
\nonumber\\
\cQ^{1,\pm}_r&=\lim_{\e\rightarrow 0}\,\sqrt{\e}\,Q^{\pm}_{r}\;,
&
\cQ^{2,\pm}_r&=\lim_{\e\rightarrow 0}\,\sqrt{\e}\,\Qbar^{\pm}_{-r}\;,
\nonumber\\
c_1&=\lim_{\e\rightarrow 0}(c-\cbar)\;,& c_2&=\lim_{\e\rightarrow 0}\,\e\,(c+\cbar)\;,
\nonumber\\
\cR_m&=\lim_{\e\rightarrow 0}\, (R_m-\Rbar_{-m})\;, & \;\cS_m&=\lim_{\e\rightarrow 0}\,\e \,(R_m+\Rbar_{-m})\;.
\end{align}
The scaled supercharges satisfy: $(\cQ^{a,\pm}_r)^\dagger=\cQ^{a,\mp}_{-r}$, with $a=1,2$.

After some simple computations, we recover the bosonic sector of the algebra \eqref{eq:algebra_gen} and some additional relations:
\begin{align}
\label{eq:N4_BMS3_algebra}
[J_m,J_n]&=(m-n)\,J_{m+n}+\frac{c_1}{12}\,m(m^2-1)\,\delta_{m+n,0}\;, \nonumber \\
[J_m,P_n]&=(m-n)\,P_{m+n}+\frac{c_2}{12}\,\,m(m^2-1)\,\delta_{m+n,0}\;,
\nonumber\\
[P_m,P_n]&=0,\quad
[P_m,\cS_n]=0, \quad [P_m,\cR_n]=-n\,\cS_{m+n}\nonumber\\
\quad [J_m,\cR_n]&=-n\,\, \cR_{m+n}\;,\quad   [J_m,\cS_n]=-n\, \cS_{m+n}
\nonumber\\
[\cR_m,\cR_n]&=\frac{c_1}{3}\,m\,\d_{m+n,0}\;,\quad [\cS_m,\cS_n]=0\;, \quad[\cR_m,\cS_n]=\frac{c_2}{3}\,m\,\d_{m+n,0}
\nonumber\\
[P_m,\cQ^{a,\pm}_r]&=0\;, \quad [J_m,\cQ^{a,\pm}_r] = \Big(\frac{m}{2}-r\Big)\,\cQ^{a,\pm}_{m+r}
\nonumber\\
[\cR_m,\cQ^{1,\pm}_r]&=\pm\cQ^{1,\pm}_{m+r}\;,\quad [\cR_m,\cQ^{2,\pm}_r]=\mp\cQ^{2,\pm}_{m+r}\;,\quad
 [\cS_m,\cQ^{a,\pm}]=0\;,
\nonumber\\
\{\cQ^{1,\pm}_r,\cQ^{1,\mp}_s\}&=\frac{1}{2}\left[P_{r+s}+\frac{1}{2}(r-s)\mathcal{S}_{r+s}+\frac{c_2}{6}(r^2-\frac{1}{4})\delta_{r+s,0}\right]
\nonumber\\
\{\cQ^{2,\pm}_r,\cQ^{2,\mp}_s\}&=\frac{1}{2}\left[P_{r+s}-\frac{1}{2}(r-s)\mathcal{S}_{r+s}+\frac{c_2}{6}(r^2-\frac{1}{4})\delta_{r+s,0}\right]
\nonumber\\
\{\cQ^{a,\pm}_r,\cQ^{b,\pm}_s\}&=0\; a \neq b.
\end{align}
where the index $a,b=1,2$.
We see that the anticommutator of each supercharge with its Hermitian conjugate closes into a linear combination of $P$ and $\cS$ plus a central term, with the coefficient of $\cS$ taking opposite signs for $a=1,2$. The anticommutator of each supercharge with itself vanishes -- as expected, given that the result has $\cR$-charge 2. The super-Poincar\'e algebra sits inside this algebra and the corresponding generators are :
\begin{equation}
J_{\pm1}, J_0,P_{\pm1},P_0, \cR_0, {\cal{Q}}^{a,\pm}_{\pm\frac{1}{2}}.
\end{equation}

\subsection{$N=4$ super-BMS$_3$ with symmetric scaling for R-currents}

In principle there is another contraction that we might want to consider. Suppose instead of asymmetrically scaling the sum and difference of the original $R$-charge, we scaled the components symmetrically. Then the last line of \eref{eq:contraction_1bis} would be replaced by:
\be
\cR_{m}=\sqrt{\epsilon}\,R_{m}\;,\quad \cS_{m}=\sqrt{\epsilon}\,\Rbar_{-m}\;.
\ee
On carrying out this scaling, we find the algebra:
\begin{align}
\label{eq:symscaling}
[J_m,J_n]&=(m-n)\,J_{m+n}+\frac{c_1}{12}\,m(m^2-1)\,\delta_{m+n,0}\;,
\nonumber\\
[J_m,P_n]&=(m-n)\,P_{m+n}+\frac{c_2}{12}\,\,m(m^2-1)\,\delta_{m+n,0}\;,
\nonumber\\
[P_m,P_n]&=0\;,\qquad\qquad\qquad [P_m,\cS_n]=0\;,\ \qquad\qquad\qquad   [P_m,\cR_n]=0\nonumber\\
[J_m,\cR_n]&=-n\,\cR_{m+n}\;,\ \qquad    [J_m,\cS_n]=-n\,\cS_{m+n}
\nonumber\\
[\cR_m,\cR_n]&=\frac{c_2}{6}\,m\,\d_{m+n,0}\;,\ \quad  [\cS_m,\cS_n]=-\frac{c_2}{6}\,m\,\delta_{m+n,0}\;,\ \quad  [\cR_m,\cS_n]=0
\nonumber\\
[P_m,\cQ^{a,\pm}_r]&=0\;, \ \qquad\qquad\quad  [J_m,\cQ^{a,\pm}_r] = \Big(\frac{m}{2}-r\Big)\,\cQ^{a,\pm}_{m+r}
\nonumber\\
[\cR_m,\cQ^{a,\pm}_r]&=0\;,\ \qquad\qquad\quad  [\cS_m,\cQ^{a,\pm}]=0\;,  \qquad  \{\cQ^{a,\pm}_r,\cQ^{b,\pm}_s\}=0\;, \; a \neq b
\nonumber\\
\{\cQ^{1,\pm}_r,\cQ^{1,\mp}_s\}&=\frac{1}{2}\Big(P_{r+s}+\tfrac1{6}\,c_2\,r^2\,\d_{r+s,0}\Big)\;,\quad
\{\cQ^{2,\pm}_r,\cQ^{2,\mp}_s\}=\frac{1}{2}\Big(P_{r+s}+\tfrac1{6}\,c_2\,r^2\,\d_{r+s,0}\Big)\;. \nonumber
\end{align}
We note that this algebra has no $R$-symmetry. Indeed, all bosonic operators commute with the $\cQ^{a,\pm}$ other than $J_m$ which just measures that its spin is $\frac12$. Likewise, $\cR$ and $\cS$ commute with everything except themselves, and with $J_n$ which again measures that their spin is 1. This algebra therefore appears trivial and is not the correct one to describe the asymptotic symmetry of flat-space extended supergravity. Thus we conclude that the correct scaling is the asymmetric one of Section (4.1) and the correct algebra is the one in \eref{eq:N4_BMS3_algebra}. For the case of $N=8$ super-BMS$_3$  that we consider in the next section, we
shall only study the asymmetric scaling for $R-$currents. 
\subsection{Generic $N=4$ super-BMS$_3$ algebra}

The algebra in eq.(\ref {eq:N4_BMS3_algebra}) obtained using the In\"on\"u-Wigner
contraction of the $N=4$ superconformal algebra has specific relations among the 
central charges appearing in the commutators.  In particular, the central
charge appearing in the $[J_m, J_n]$ is related to the central charge
appearing in the $[\mathcal{R}_m, \mathcal{R}_n]$ commutator.
However, this condition turns out to be too restrictive. To find the most general allowed central charges, one should use the
Jacobi identies.  The central charges of the algebra before contraction were chosen to satisfy the Jacobi identities of that algebra, therefore the same is true of the result after contraction. However, since the contracted algebra has fewer generators, it is logically  possible that there may be greater freedom in choosing the central extensions of the contracted algebra while continuing to satisfy the Jacobi identity of this algebra.

In fact, this is precisely the case for the $N=4$ super-BMS algebra
given in eq.\eqref {eq:N4_BMS3_algebra}.  Using the commutation relations among the algebra
elements and keeping the central extensions arbitrary subject to the
condition that they satisfy the Jacobi identity, we find that the
central term appearing the the $[\mathcal{R}_m, \mathcal{R}_n]$
commutator is not related to that in the $[J_m, J_n]$ commutator. This is because these two were originally related to each other through the central charge of the supersymmetry algebra. But after contraction, the supersymmetry algebra produces the ${\cal S}$ generator on the right-hand-side instead of the R-symmetry generator ${\cal R}$. Thus  a relation no longer holds between the Virasoro and R-symmetry central charges, and the general $N=4$ super-BMS algebra therefore has independent central
terms in the $[\mathcal{R}_m, \mathcal{R}_n]$ and $[J_m, J_n]$
commutators. 

From now on we will consider the $N=4$ super-BMS algebra with these central terms being independent. Consistency of this algebra is confirmed by the free field realization of this algebra.  As we will see shortly, this realization naturally produces different
central charges for these two commutators.  The central extension in
the $[\mathcal{R}_m, \mathcal{R}_n]$ commutator can, if desired, be varied by adding more free fields to the base system.

\subsection{Free-Field Realization of this Algebra}

As in the previous case, we present the free-field realizations of this most generic $N=4$ extended super BMS$_3$ algebra. Accordingly, the central charge in the $\mathcal{R}-\mathcal{R}$ OPE is now labelled $c_3$. The free-field realization supports 
our claim of the last section, that it provides us independent central terms for $J-J$ and $\cR-\cR$ commutators. Below, we first rewrite the algebra in terms of the OPE's of various fields:
\be
   \label{eq:11b}
     \begin{split}
T(z)T(w)&\sim\frac{\frac{c_1}{2}}{(z-w)^4}+\frac{2T(w)}{(z-w)^2}+\frac{\partial T}{z-w}, \quad
\mathcal{R}(z)\mathcal{R}(w)\sim\frac{c_3}{3}\frac{1}{(z-w)^2},\\
T(z)P(w)&\sim\frac{\frac{c_2}{2}}{(z-w)^4}+\frac{2P(w)}{(z-w)^2}+\frac{\partial P}{z-w},\quad
\mathcal{S}(z)\mathcal{R}(w)\sim\frac{c_2}{3}\frac{1}{(z-w)^2} ,\\
T(z)S(w)&\sim\frac{\mathcal{S}(w)}{(z-w)^2}+\frac{\partial S(w)}{z-w}, \quad
T(z)R(w)\sim\frac{R(w)}{(z-w)^2}+\frac{\partial R(w)}{z-w}, \\
P(z)\mathcal{R}(w)&\sim\frac{\mathcal{S}(w)}{(z-w)^2}+\frac{\partial \mathcal{S}(w)}{z-w}, \quad
P(z)\mathcal{Q}^{a,\pm}\sim 0, \\
T(z)\mathcal{Q}^{a,\pm}(w)&\sim\frac{3}{2}\frac{\mathcal{Q}^{a,\pm}(w)}{(z-w)^2}+\frac{\partial\mathcal{Q}^{a,\pm}}{z-w},\quad
\mathcal{S}(z)\mathcal{S}(w)\sim0, \quad
P(z)\mathcal{S}(w)\sim 0,\\
\mathcal{S}(z)\mathcal{Q}^{a\pm}(w)&\sim 0 ,\quad
\mathcal{R}(z)\mathcal{Q}^{1,\pm}(w) \sim\pm\frac{\mathcal{Q}^{1,\pm}}{z-w}, \quad
\mathcal{R}(z)\mathcal{Q}^{2,\pm}(w)\sim\mp\frac{\mathcal{Q}^{2,\pm}}{z-w},\\
\mathcal{Q}^{1,\pm}(z)\mathcal{Q}^{1,\mp}(w)&\sim\frac{1}{2}\left[\frac{P(w)}{z-w}+\frac{1}{2}\left\{\frac{2\mathcal{S}(w)}{(z-w)^2}+\frac{\partial \mathcal{S}}{z-w}\right\}+\frac{c_2}{3}\frac{1}{(z-w)^3}\right] ,\\
\mathcal{Q}^{2,\pm}(z)\mathcal{Q}^{2,\mp}(w)&\sim\frac{1}{2}\left[\frac{P(w)}{z-w}-\frac{1}{2}\left\{\frac{2\mathcal{S}(w)}{(z-w)^2}+\frac{\partial \mathcal{S}}{z-w}\right\}+\frac{c_2}{3}\frac{1}{(z-w)^3}\right] \
   \end{split}
\ee

Let us now give a free-field representation of this algebra. For this we need the fields $\left(\beta_2,\gamma_{-1}\right)$, $\left(\beta_1,\gamma_0\right)$ and 4 pairs of fermionic fields $\left(\chb^{a,\alpha},\chc^{a,\alpha}\right)$ 
where, $a=1,2$ and $\alpha=\pm$.  With these, we define,
\be
\label{eq.4.33}
\begin{split}
 &T_{(2,-1)} = -2\beta_2\partial\gamma_{-1} -
    \gamma_{-1}\partial\beta_2 , ~~~
    T_{(1,0)} = -\beta_1\partial\gamma_0,\\
    &T^{a,\alpha}_{(3/2,-1/2)} = -\frac32 \chb^{a,\alpha}\partial \chc^{a,\alpha} + \frac12 \chc^{a,\alpha}\partial \chb^{a,\alpha} .
\end{split}
\ee
Finally, we identify various $N=4$ fields as follows:
\be
   \label{eq:11c}
     \begin{split}
T &=T_{(2,-1)} + T_{(1,0)}+\sum_{i=1}^2\sum_{\alpha=1}^2
    T^{a,\alpha}_{(3/2,-1/2)}- \lambda \partial^3\gamma_{-1}^ 1\ ,\qquad  P=\beta_2,\\
\mathcal{R}&=\partial\gamma_0+\kappa\partial\beta_1\gamma_{-1}+\kappa\beta_1\partial\gamma_{-1}+\sum_{a=1}^2 (\chb^{a,\alpha}(\sigma_3) _{\alpha\beta}\,\chc^{a,\beta})\ , \qquad \mathcal{S}=-\kappa\beta_1\\
\mathcal{Q}^{a,\alpha}&=\frac{1}{2}\left(\chb^{a,\alpha}+\beta_2(\sigma_1)^\alpha_\beta\,\chc^{a,\beta}+\rho\partial\beta_1(i\sigma_2)^\alpha_\beta\,\chc^{a,\beta}+2\rho\beta_1(i\sigma_2)^\alpha_\beta\,\partial\chc^{a,\beta}+\eta(\sigma_1)^\alpha_\beta\,\partial^2\chc^{a,\beta}\right)
 \end{split}
\ee
This set of fields correctly reproduce the $N=4$ super BMS$_3$ algebra, 
where we need to identify $(\lambda = \frac{c_2}{12}, \kappa=\frac{c_2}{3}, \rho=\frac{c_2}{6}, \eta=\frac{c_2}{6})$. We get $c_1$ and 
$c_3$, the central charges in the $T-T$ and $\mathcal{R}-\mathcal{R}$ OPEs of the most generic BMS$_3$ algebra respectively, fixed as $c_1=-32$ and $c_3=12$.
These two can be made arbitrary by adding independent free fields to our system.


\section{N$=8$ SBMS$_3$}
\label{sec:5}
Finally, we look at systems with eight supercharges, i.e.
N$=8$ SBMS$_3$. This can be obtained by contracting two copies of the N$=4$ superconformal algebra. In this section, we shall consider two copies of small  of N$=4$ superconformal algebra.  The small $N = 4$ superconformal algebra is
generated by bosonic currents $T, J^i$
 with $ (i = 1, 2, 3)$ and fermionic currents $G^{a,\alpha}$
 with
$(a,\alpha = 1, 2)$.  The central charge is related to the level of the $SU(2)$ currents. We present the algebra in terms of the modes $(L_m, T^{i}_m,Q^{a,\alpha}_r)$ for holomorphic currents $(T, J^i,G^{a,\alpha})$ respectively. Similarly the antiholomorphic sector will be represented by  $\bar L_m, \bar T^{,i}_m,\bar Q^{a,\alpha}_r$ modes.
The algebra reads as\cite{Ito:1998vd}:
\begin{align}
\label{eq:N4_SCA_algebra}
[L_m,L_n]&=(m-n)L_{m+n}+\frac{c}{12}m(m^2-1)\delta_{m+n,0} , \quad [L_m,T^{i}_n]=-nT^{i}_{m+n} , \nonumber \\
[L_m,Q^{a,\alpha}_r]&=\left(\frac{m}{2}-r\right)Q^{a,\alpha}_{m+r} , \quad
[T^{i}_m,T^{j}_n]=i\epsilon^{ijk}T^{k}_{m+n}+\frac{c}{12}m\delta^{ij}\delta_{m+n,0},\nonumber \\
[T^{i}_m,Q^{a,1}_r]&=-\frac{1}{2}\left(\sigma^i\right)^a_bQ^{b,1}_{m+r}, \quad
[T^{i}_m,Q^{a,2}_r]=\frac{1}{2}\left(\bar{\sigma}^i\right)^a_bQ^{b,2}_{m+r}\nonumber \\
\{Q^{a,+}_r,Q^{b,-}_s\}&=\left[\delta^{ab}L_{r+s}-(r-s)(\sigma^i)_{ab}T^{i}_{r+s}+\frac{c}{6}(r^2-\frac{1}{4})\delta^{ab}\delta_{r+s,0}\right] . 
\end{align}
together with the anti-holomorphic counterpart with an independent central charge $\cbar$. Here, $\bar{\sigma}^i_{ab}= \sigma^i_{ba}$ and ${\sigma}^i$ are the three Pauli matrices. In the next section, we present the flat space limit of this algebra. 

\subsection{N=8 SBMS$_3$ with asymmetric scaling for the R-currents}

In the previous section, for N=4 SBMS$_3$, we have already seen that only the asymmetric scaling for the R-currents provide a sensible
algebra with non-trivial R-symmetry. Hence, in this section, we shall only present the results with asymmetric scaling. The symmetric scaling is trivial as the earlier cases.
Let us now define the following operators by contracting the two copies of N$=4$ SCA.
\begin{align}
P_m&=\lim_{\epsilon\to 0}\epsilon\left(L_m+\bar L_{-m}\right)\;,\qquad &
J_m=\lim_{\epsilon\to 0}\left(L_m-\bar L_{-m}\right)\nonumber \\
c_1&=\lim_{\epsilon\to 0}\left(c-\cbar\right) \;,\qquad &
c_2=\lim_{\epsilon\to 0}\epsilon \left(c+\cbar\right)   \\
\mathcal{R}^i_m&=\lim_{\epsilon\to 0}\left(T^{i}_m+\bar T^{i}_{-m}\right)\;,\qquad &
\mathcal{S}^i_m=\lim_{\epsilon\to 0}\epsilon\left(T^{i}_m-\bar T^{i}_{-m}\right)
\nonumber \\
\mathcal{Q}^{1,a,\alpha}_r&=\lim_{\epsilon\to 0}\sqrt{\epsilon}Q^{a,\alpha}_{ r} \;,\qquad &
\mathcal{ Q}^{2,a,\alpha}_r=\lim_{\epsilon\to 0}\sqrt{\epsilon} \bar Q^{a,\alpha}_{- r}\, ,
\nonumber 
\end{align}
where, $a=1,2$ and $\alpha= \pm$.  We then get the commutation relations, 
\begin{align}
\label{eq:N4,4_SBMS3}
&[J_m,J_n]=(m-n)\,J_{m+n}+\frac{c_1}{12}\,m(m^2-1)\,\delta_{m+n,0}\;, \qquad\qquad\qquad\quad \left[\mathcal{S}^i_m,\mathcal{S}^j_n\right]=0\;,
\nonumber\\
&[J_m,P_n]=(m-n)\,P_{m+n}+\frac{c_2}{12}\,\,m(m^2-1)\,\delta_{m+n,0}\;, \qquad\qquad\qquad\quad [P_m,P_n]=0\;,
\nonumber\\
&\left[P_m,\mathcal{R}^i_n\right]=-n\mathcal{S}^i_{m+n}\;, \quad \left[P_m,\mathcal{S}^i_n\right]=0\;, \quad  \left[J_m,\mathcal{R}^i_n\right]=-n\mathcal{R}^i_{m+n}\;, \quad  \left[J_m,\mathcal{S}^i_n\right]=-n\mathcal{S}^i_{m+n} \nonumber \\
&\left[\mathcal{R}^i_m,\mathcal{R}^j_n\right]=i\epsilon^{ijk}\mathcal{R}^k_{m+n}+\frac{c_1}{12}m\delta^{ij}\delta_{m+n,0}\;,\quad
[\cR^i_m,\cS^j_n]={\rm i}\,\e^{ijk}\,\cS^k_{m+n}+\frac{c_2}{12}m\delta^{ij}\delta_{m+n,0}
\nonumber\\
&\left[P_m,\mathcal{Q}^{A,a,\alpha}_r\right]=0\;, \quad
\left[J_m,\mathcal{Q}^{A,a,\alpha}_r\right]=\left(\frac{m}{2}-r\right)\mathcal{Q}^{A,a,\alpha}_{m+r}\;,\qquad 
 \left[\mathcal{S}^i_m,\mathcal{Q}^{A,a,\alpha}_r\right]=0 \\
 & \left[\mathcal{R}^i_m,\mathcal{Q}^{A,a,1}_r\right]=-\frac{1}{2}\left(\sigma^i\right)^a_b\mathcal{Q}^{A,b,1}_{m+r}\;, \qquad 
 \left[\mathcal{R}^i_m,\mathcal{Q}^{A,a,2}_r\right]=\frac{1}{2}\left(\bar {\sigma}^i\right)^a_b\mathcal{Q}^{A,b,2}_{m+r}\nonumber\\
&\{\mathcal{Q}^{A,a,+}_r,\mathcal{Q}^{A,b,-}_s\}=\frac{1}{4}\left[1+(-1)^{A+B}\right]
\left[\delta^{ab}P_{r+s}-(r-s)(\sigma^i)_{ab}\mathcal{S}^{i}_{r+s}+\frac{c_2}{6}(r^2-\frac{1}{4})\delta^{ab}\delta_{r+s,0}\right]\nonumber \
\end{align}
where $A,B=1,2$.  This is the N=8 SBMS$_3$ algebra, with $\mathcal{R}^i_m$ being the modes of the R-symmetry generators. Like the $N=4$ case, even this algebra is too restrictive in its central extension. Applying Jacobi identity, we find as before that the most generic version of this $N=8$ super BMS$_3$ algebra has an independent central term $c_3$in the $\cR-\cR$ commutator. Notice that, as in the earlier cases, we can clearly identify the super-Poincar\'e algebra as the sub algebra of this new algebra consisting of the following generators :
\begin{equation}
J_{\pm1}, J_0,P_{\pm1},P_0, \cR^i_0, {\cal{Q}}^{A,a,\pm}_{\pm\frac{1}{2}}.
\end{equation}

\subsection{Free-Field Realization of this Algebra}
Finally, we present the free-field realization of the most generic three dimensional $N=8$ super BMS$_3$ algebra. For this, 
first we express the algebra in terms of operator product expansions:
\be
    \label{eq:17}
  \begin{split}
&T(z)T(w)\sim\frac{\frac{c_1}{2}}{(z-w)^4}+\frac{2T(w)}{(z-w)^2}+\frac{\partial T}{z-w}, \quad
T(z)P(w)\sim\frac{\frac{c_2}{2}}{(z-w)^4}+\frac{2P(w)}{(z-w)^2}+\frac{\partial P}{z-w},\\
&\mathcal{R}^i(z)\mathcal{R}^j(w)\sim\frac{i\epsilon^{ijk}\mathcal{R}^k}{z-w}+\frac{c_3}{12}\frac{\delta^{ij}}{(z-w)^2},\qquad 
\mathcal{R}^i(z)\mathcal{S}^j(w)\sim\frac{i\epsilon^{ijk}\mathcal{S}^k}{z-w}+\frac{c_2}{12}\frac{\delta^{ij}}{(z-w)^2} ,\\
&T(z)\mathcal{S}^i(w)\sim\frac{\mathcal{S}^i(w)}{(z-w)^2}+\frac{\partial \mathcal{S}^i(w)}{z-w} , \quad
T(z)\mathcal{R}^i(w)\sim\frac{\mathcal{R}^i(w)}{(z-w)^2}+\frac{\partial \mathcal{R}^i(w)}{z-w}, \\
&P(z)\mathcal{R}^j(w)\sim\frac{\mathcal{S}^j(w)}{(z-s)^2}+\frac{\partial \mathcal{S}^j(w)}{z-w},\qquad
T(z)\mathcal{Q}^{A,a,\pm}(w)\sim\frac{3}{2}\frac{\mathcal{Q}^{A,a,\pm}(w)}{(z-w)^2}+\frac{\partial\mathcal{Q}^{A,a,\pm}(w)}{z-w},\\
&\mathcal{S}^i(z)\mathcal{Q}^{A,a,\pm}(w)\sim 0, \quad \mathcal{S}^i(z)\mathcal{S}^j(w)\sim0, \quad P(z)\mathcal{S}^i(w)\sim 0, \quad
P(z)\mathcal{Q}^{A,a,\pm}\sim 0\\
&\mathcal{R}^i(z)\mathcal{Q}^{A,a,+}(w)\sim -\frac{1}{2}\frac{\left(\sigma^i\right)^a_b\mathcal{Q}^{A,b,+}(w)}{z-w},, \quad \mathcal{R}^i(z)\mathcal{Q}^{A,a,-}(w)\sim \frac{1}{2}\frac{\left(\bar{\sigma}^i\right)^a_b\mathcal{Q}^{A,b,-}(w)}{z-w}, \quad\\
&\mathcal{Q}^{A,a,\pm}(z)\mathcal{Q}^{B,b,\mp}(w)\sim\frac{1}{2}\delta^{AB}\left[\frac{\delta^{ab} P(w)}{z-w}-(\sigma^i)_{ab}\left(\frac{\mathcal{S}^i(w)}{(z-w)^2}+\frac{\partial\mathcal{S}^i(w)}{z-w}\right)+\frac{c_2}{3}\frac{1}{(z-w)^3}\right]\
\end{split}
\ee



For a free-field realization of  $\mathcal{N}=8$ BMS, we introduce one pair of conjugate bosonic ghost-fields $\left(\beta_2,\gamma_{-1}\right)$, three pairs of bosonic ghost fields $\left(\beta_1^i,\gamma_{0}^i\right)$, {\em i.e.}, $i=1,2,3$ and eight pairs of fermionic ghost fields $\left(\chb^{A,a,\alpha},\chc^{A,a,\alpha}\right)$
where both $A$ and $a$ take the values $1$ and $2$, and $\alpha=\pm$.  Finally, using eqn. (\ref{eq.4.33}), the $N=8$-fields can be expressed as follows:
\begin{equation}
  \begin{split}
    T &=T_{(2,-1)} +  \sum_{i=1}^3T^i_{(1,0)} +\sum_{A=1}^a \sum_{a=1}^2 \sum_{\alpha=\pm}
    T^{A,a,\alpha}_{(3/2,-1/2)} - \lambda \partial^3\gamma_{-1}\ , \quad
    P = \beta_2\ , \\
    \mathcal{R}^i &= \partial\gamma_0^i+\kappa (\beta_1^i\partial\gamma_{-1}
      + \partial\beta_1^i\gamma_{-1})+
      i\epsilon^{ijk}\gamma_0^j\beta_1^k  \\ &\hspace{4mm}+ \frac12 \chc^{A,a,\alpha}(\sigma^i)_{ab}(\sigma_3)_{\alpha\beta}\chb^{A,b,\beta}\,
    , \qquad \mathcal{S}^i = -\kappa\beta^i_1\ ,\\
\mathcal{Q}^{A,a,+} &= \frac{1}{2}\left[\chb^{A,a,+}+\beta_2(\sigma_1)^+_{~-} \,\chc^{A,a,-}+\rho\,(\sigma^j)_{ae}\left\{\partial\beta^i_1(i\sigma_2)^+_{~-}\,\chc^{A,e,-}\right.\right.\\ &\hspace{4mm}\left.\left.+2\beta^i_1(i\sigma_2)^+_{~-}\,\partial \chc^{A,e,-} \right\}+\eta\,(\sigma_1)^+_{~-}\,\partial^2\chc^{A,a,-}\right], \\
\mathcal{Q}^{A,a,-} &= \frac{1}{2}\left[\chb^{A,a,-}+\beta_2(\sigma_1)^-_{~+}\, \chc^{A,a,+}+\rho\,(\bar\sigma^j)_{ae}\left\{\partial\beta^i_1(i\sigma_2)^-_{~+}\,\chc^{A,e,+}\right.\right.\\ &\hspace{4mm}\left.\left. +2\beta^i_1(i\sigma_2)^-_{~+}\,\partial \chc^{A,e,+} \right\}+\eta\,(\sigma_1)^-_{~+}\,\partial^2\chc^{A,a,+}\right], 
  \end{split}
\end{equation}
where, repeated indices in the expressions of $\mathcal{R}^i$ and $\mathcal{Q}^{A,a,\alpha}$ are summed.
The parameters $\lambda$, $\eta$, $\kappa$ and $\rho$ are not independent. 
We can write all these parameters in terms of $c_2$ as follows: $\lambda=\frac{c_2}{12} , \eta=\frac{c_2}{6}$, $\rho=-\frac{c_2}{12}$ and $\kappa=\frac{c_2}{12}$. We find $c_1=-88$ and $c_3=24$.

\section{Summary} 
\label{sec:6}

\begin{table}
\centering
\begin{tabular}{@{\extracolsep{\fill}}|P{2.5cm}|P{4.5cm}|P{5.5cm}|}
\hline
Super BMS$_3$  & Global super-Poincare' generators &  Free fields 
   \\ 
 \hline 
 \noalign{\smallskip}
N=1  & $J_{\pm 1}, J_0,P_{\pm 1},P_0, {\cal{Q}}_{\pm\frac{1}{2}}$ & $(\beta_2 - \gamma_{-1})$ and $(\chb-\chc)$
  \\[1mm] 
 \hline  
 \noalign{\smallskip}

N=2  & $J_{\pm 1}, J_0,P_{\pm 1},P_0, {\cal{Q}}^{a}_{\pm\frac{1}{2}}$ & $(\beta_2 - \gamma_{-1})$ and $(\chb^a-\chc^a)$
  \\[1mm] 
 \hline  
 \noalign{\smallskip}
N=4  & $J_{\pm 1}, J_0,P_{\pm 1},P_0, \cR_0, {\cal{Q}}^{a,\alpha}_{\pm\frac{1}{2}}$ & $(\beta_2 - \gamma_{-1}, \quad \beta_1-\gamma_0)$ and $(\chb^{a, \alpha}-\chc^{a, \alpha})$
  \\[1mm] \hline\noalign{\smallskip}
N=8  & $J_{\pm 1}, J_0,P_{\pm 1},P_0, \cR^i_0, {\cal{Q}}^{A,a,\alpha}_{\pm\frac{1}{2}}$ & $(\beta_2 - \gamma_{-1}, \quad \beta^i_1-\gamma^i_0)$ and $(\chb^{A,a, \alpha}-
\chc^{A,a, \alpha})$
  \\[1mm] \hline\noalign{\smallskip}
\end{tabular}
\vskip 2mm
\caption{\label{table:1} {\footnotesize {The global super-Poincare generators and the free fields that represent the algebra are listed. The indices $(A,a)$ run over $(1,2)$, $\alpha$ runs over $(\pm)$ and $i$ runs over $(1,2,3)$.}}}
\end{table}
In this paper we studied BMS$_3$ algebras with $N=2,4$ and $8$ extended supersymmetry.  They were constructed by carrying out the In\"on\"u-Wigner contraction of appropriate superconformal algebras. We have also provided free field representations for all these algebras, on the lines of those provided for BMS$_3$ and its $N=1$ supersymmetric extension in \cite{Banerjee:2015kcx}. We found that although in principle there are various ways in which extended superconformal algebras can be contracted, imposing the physical requirements that the super-Poincar\'e algebra appears correctly and that there is a residual R-symmetry that rotates supercharges into each other severely constrains the possible contractions.  We found that only the asymmetric scaling of the R-symmetry leads to a sensible extended super-BMS$_3$ algebra, where the R-symmetry rotates the supercharges into each other. Symmetric scaling of the R-symmetry generators instead gives an algebra where the residual U(1) symmetry commutes with supersymmetry.  

We also explicitly constructed free field realizations of all the extended super-BMS$_3$ algebras using the $\beta-\gamma$ and $\chb-\chc$ ghost systems.  We found that the free field representation for $N=2$ super-BMS$_3$ case is a straightforward generalization of the one for $N=1$ super-BMS$_3$, but the $N=4$ and $N=8$ cases involve novel features due to the presence of R-symmetry.  In the case of $N=4$ we have U(1) R-symmetry and for $N=8$ we have SU(2) R-symmetry.  The latter case, due to the non-abelian R-symmetry,  is quite tightly constrained. The results are presented in the table \ref{table:1} above.

In the $N=4$ and $N=8$ case we encountered an interesting phenomenon, which was in fact uncovered by the free field realization of these algebras.  As mentioned earlier one way of obtaining these algebras is contraction of the superconformal algebra in two dimensions.  However, the original algebra restricts the central terms that appear in the contracted algebra.  Closure of the contracted algebra, however, allows more freedom in the central terms.  We showed that the simplest free field realization for these algebras indeed give rise to distinct central terms, i.e. the central charge appearing in the R-symmetry commutators is decoupled from the one appearing in the commutator of the super-rotation generators.  Free field realizations with equal central charges can also be obtained, but in that case they are non-minimal, in the sense that we require additional fields to adjust the central term.

Another outcome of this investigation is the clear distinction between the super-BMS$_3$ and super-GCA$_2$ algebras.  Although the bosonic GCA$_2$ and BMS$_3$ are isomorphic, this correspondence does not extend to the super-BMS algebras.  We studied a variety of contractions of (1,1) superconformal algebras and in each case we found that in order to relate the super-BMS$_3$ to super-GCA$_2$ we either have to give up on the Poincar\'e subalgebra or make the supercharges complex.  In either case, there is no clear physical interpretation that can justify the identification. 
\vspace{1cm}

\noindent 
{\bf Acknowledgements}\\
\noindent
IL would like to thank Bernard de Wit, Daniel Butter and Valentin Reys for useful discussions. TN would like to thank HRI, Allahabad for
hospitality during the course of the work. Our work is partially supported by the following
Government of India Fellowships/Grants: NB and IL by a Ramanujan Fellowship, DST; DPJ by
DAE XII-plan grant 12-R \&
D-HRI-5.02-0303; SM by a J.C. Bose Fellowship, DST; and TN
by a UGC Fellowship. We thank the people of India for their generous support to the
basic sciences.

\appendix
\section{$N=2$ GCA$_2$ and BMS$_3$ algebras}
The GCA$_2$ algebra has been studied in detail in the literature, see for instance \cite{Bagchi:2009my}. We will briefly review the relevant results and comment on the isomorphism with the BMS$_3$ algebra, which at the supersymmetric level gets lifted. One starts from linear combinations of holomorphic/anti-holomorphic Virasoro generators which maintain the mode number\cite{Bagchi:2010zz}, and scales asymmetrically as follows:
\begin{equation}
\label{eq:contraction_gen_Ips}
P_m=\lim_{\e\rightarrow 0}\e\,(L_m\mp \bar L_m) \;,\qquad J_m=\lim_{\e\rightarrow 0}(L_m \pm \bar L_m) \,.
\end{equation}
The commutators turn out to be:
\begin{align}
\label{eq:bosonic_algebra_NM}
[P_m,P_n]&=0
\nonumber\\
[P_m,J_n]&=(m-n)\,\lim_{\e\rightarrow 0}\,\epsilon\, (L_{m+n}-\Lbar_{m+n})+\dots
\nonumber\\
[J_m,J_n]&=(m-n)\,\lim_{\e\rightarrow 0}\, (L_{m+n}+\Lbar_{m+n})+\dots
\end{align}
Thus the algebra closes, and if we fix the signs in \eqref{eq:contraction_gen_Ips} to be minus in the definition of $P_m$ and plus sign in the definition of $J_m$ then we recover the BMS$_3$ algebra.

An N=2 generalisation of the GCA algebra was presented in Ref.\cite{Mandal:2010gx}. It involves an asymmetric scaling of the form:
\be
\cQ^1_r=\lim_{\e\rightarrow 0}\e\,(Q_r\mp \bar Q_r) \;,\qquad \cQ^2_r=\lim_{\e\rightarrow 0}(Q_r \pm \bar Q_r)
\ee
The choice of upper/lower signs is immaterial as it simply corresponds to a sign change for $\Qbar$. The resulting algebra is:
\begin{align}
\label{eq:fermionic_algebra_NM_case1}
[P_m,\cQ^1_r]&=0\;,\qquad \qquad\qquad\qquad\qquad \qquad\qquad [P_m,\cQ^2_r]=\Big(\frac{m}2-r\Big)\cQ^1_{m+r}
\nonumber\\
[J_m,\cQ^1_r]&=\Big(\frac{m}2-r\Big)\cQ^1_{m+r}\;,\qquad\qquad \qquad\qquad[J_m,\cQ^2_r]=\Big(\frac{m}2-r\Big)\cQ^2_{m+r}
\nonumber\\
\{\cQ^1_r,\cQ^1_s\}&=0\;,\quad\quad \{\cQ^1_r,\cQ^2_s\}=2\,P_{r+s}+\dots \;,\quad \{\cQ^2_r,\cQ^2_s\}=2\,J_{r+s}+\dots
\end{align}
We can now examine whether this algebra is isomorphic to the N=2 super-BMS algebra in Eq.(\ref{eq:BMS_ferm_alg}). Clearly it is not: the supercharge anti-commutators can be diagonalised to find that one of them has a negative right-hand-side. This shows that the $N=2$ super-GCA of Ref.\cite{Mandal:2010gx} is not equivalent to the $N=2$ super-BMS$_3$ algebra. Thus the BMS/GCA correspondence does not hold in the supersymmetric case.\\
In \cite{Lodato:2016}, this asymptotic superalgebra is studied in detail and proven to arise from a `twisted' novel  supersymmetric  theory in 3 dimensions.\\

One may try to scale the super-generators symmetrically:
\be
\cQ_r^+=\lim_{\epsilon\to 0}\sqrt{\epsilon}Q_r,\qquad
\cQ_r^-=\lim_{\epsilon\to 0}\sqrt{\epsilon}\Qbar_r
\ee
This is similar (except for the fact that mode number is preserved) to the symmetric scaling used in super-BMS, but the bosonic generators are scaled according to GCA and the resulting algebra therefore contains:
\begin{align}
\label{eq:fermionic_algebra_NM_case2}
[P_m,\cQ^\pm_r]&=0\;,\qquad\qquad\qquad [J_m,\cQ^\pm_r]=\Big(\frac{m}2-r\Big)\cQ^\pm_{m+r}
\nonumber\\
\{\cQ^\pm_r,\cQ^\pm_s\}&=\pm( P_{r+s}+\dots) \;,\quad \{\cQ^+_r,\cQ^-_s\}=0
\end{align}
We see that the RHS has a negative sign in front of $P$ for one of the generators. Therefore this also cannot be identified with the super-BMS$_3$ algebra. 

One might be tempted to redress this by inserting a factor of $i$:
\be
\label{eq:non-Hermitian}
 \cQ^+_r=\lim_{\e\rightarrow 0}\,\sqrt{\e}\,Q_r\;,\qquad\qquad \cQ^-_r=\lim_{\e\rightarrow 0}\,{\rm i}\,\sqrt{\e}\,\bar Q_r 
\ee
but unfortunately this implies that the hermiticity condition on $\cQ^-$ is violated. Thus we really get nothing new.

Finally, let us comment on a proposal in Ref.\cite{Bagchi:2016yyf}. These authors propose to recover the SGCA \eqref{eq:fermionic_algebra_NM_case1} by defining an ``inhomogeneous scaling'' for the supercharges: 
\be
\cQ^1_r=\lim_{\e\rightarrow 0}\e\,(Q^+_n\pm {\rm i} Q^-_{-n})\;,\qquad  \cQ^2_r=\lim_{\e\rightarrow 0}(Q^+_n \mp {\rm  i} Q^-_{-n}) \;.
\ee
Unfortunately this suffers from an analogous defect to Eq.(\ref{eq:non-Hermitian}) above, namely the supercharges do not satisfy ${\cQ^i_r}^\dagger=\cQ^i_{-r}$. Instead one finds that ${\cQ^2_r}^\dagger\sim \frac{1}{\epsilon}\cQ^1_{-r}$ and ${\cQ^1_r}^\dagger\sim \epsilon\cQ^2_{-r}$ so the hermiticity properties are incompatible with the scaling.

\section{Inequivalent $N=2$ super-BMS$_3$ from (1,1) and (2,0) Virasoro algebras}\label{sec:ineq-n=2-super}
There is another way of constructing $N=2$ super BMS$_3$ algebra than the one presented in the main draft.
To obtain this algebra we need to consider only one sector of super-conformal algebra, which for definiteness  can be taken to be the holomorphic sector, for the supercharges and the R-symmetry generators. The R-generators can be scaled in two ways:
$\cR_m=\lim_{\e\rightarrow 0}\,\e \,R_m$ or $\cS_m=\lim_{\e\rightarrow 0}\,R_m$, while the remaining generators are scaled as usual.

Let's consider the first scaling. The commutation relations \eqref{eq:N4_BMS3_algebra} will still be valid except that there is no generator corresponding to $\cS$, so we find the algebra obtained by setting $\cS=0$ there:
 \begin{align}
\label{eq:N2_20_a_BMS3_algebra}
[P_m,\cR_n]&=0\;,\quad \quad [J_m,\cR_n]=-n\,\cR_{m+n}\;,\qquad [\cR_m,\cR_n]=0
\nonumber\\
 [P_m,\cQ^{+\,,i}_r]&=0\;, \qquad [J_m,\cQ^{+\,,i}_r]=\Big(\frac{m}{2}-r\Big)\,\cQ^{+\,,i}_{m+r}
\nonumber\\
[\cR_m,\cQ^{+\,,i}_r]&=0\;,\qquad \{\cQ^{+\,1}_r,\cQ^{+\,,2}_s\}= \tfrac1{2}\,P_{r+s}+ \tfrac14\,(r-s)\,\cR_{r+s}+\frac{c_2}{12}\,r^2\,\d_{r+s,0} 
\end{align}
This is a consistent algebra, which differs from the $N=2$ BMS$_3$ algebra we found before because of the presence of the $\cR$ generator. However this is not an R-symmetry since it does not rotate the supercharges but instead commutes with them. 

Next consider the second scaling, i.e. the generator $\cS_m=\lim_{\e\rightarrow 0}\, R_m$. The algebra will now look like:
\begin{align}
\label{eq:N2_20_b_BMS3_algebra}
[P_m,\cS_n]&=0\;,\qquad\qquad\qquad\quad  [J_m,\cS_n]=-n\,\cS_{m+n}\;,\qquad\qquad [\cS_m,\cS_n]=\frac{c_1}{3}\,m\,\d_{m+n,0}\;,
\nonumber\\
 [P_m,\cQ^{+\,,i}_r]&=0\;, \qquad \qquad\qquad\;[J_m,\cQ^{+\,,i}_r]=\Big(\frac{m}{2}-r\Big)\,\cQ^{+\,,i}_{m+r}
\nonumber\\
 [\cS_m,\cQ^{+\,,1}_r]&=\,\cQ^{+\,1}_{m+r}\;, \qquad\qquad\; [\cS_m,\cQ^{+\,,2}_r]=-\,\cQ^{+\,2}_{m+r}
\nonumber\\
\{\cQ^{+\,1}_r,\cQ^{+\,,2}_s\}&= \tfrac1{2}\,P_{r+s}+\frac{c_2}{12}\,r^2\,\d_{r+s,0} 
\end{align}
This time the generator $\cS$ can be considered an R-symmetry generator since it rotates the supercharges, but it does not appear on the RHS of the anticommutator of two $\cQ$'s. Hence, although this seems to be a valid alternate super BMS$_3$ algebra, it is not as rich as the one presented in the main draft. Similar behavior will hold for higher extended algebras.

\end{document}